\documentclass[aip,reprint]{revtex4-1}
\usepackage{graphicx}% Include figure files
\usepackage{dcolumn}% Align table columns on decimal point
\usepackage{bm}% bold math
\usepackage{color}% color text

\newif\ifgraph

\graphtrue             %
\begin{document}
\title{\color{blue}
Memory effects and active Brownian diffusion}

\author{Pulak K. Ghosh}
\affiliation{Department of Chemistry, Presidency University, Kolkata 700073, India}

\author{Yunyun Li}
\email{yunyunli@tongji.edu.cn}
\affiliation{Center for Phononics and Thermal Energy Science, Tongji University, Shanghai 200092, People's Republic of China}

 \author{Giampiero Marchegiani}
 \affiliation{Dipartimento di Fisica, Universit\`{a} di Camerino, I-62032 Camerino, Italy}

\author{Fabio Marchesoni}
\affiliation{Center for Phononics and Thermal Energy Science, Tongji University, Shanghai 200092, People's Republic of China}
\affiliation{Dipartimento di Fisica, Universit\`{a} di Camerino, I-62032 Camerino, Italy}

\date{today}
\begin{abstract}
A self-propelled artificial microswimmer is often modeled as a
ballistic Brownian particle moving with constant speed aligned along
one of its axis, but changing direction due to random collisions with
the environment. Similarly to thermal noise, its angular
randomization is described as a memoryless stochastic process. Here,
we speculate that finite-time correlations in the orientational
dynamics can affect the swimmer's diffusivity. To this purpose we
propose and solve two alternative models. In the first one we simply
assume that the environmental fluctuations governing the swimmer's
propulsion are exponentially correlated in time, whereas in the
second one we account for possible damped fluctuations of the
propulsion velocity around the swimmer's axis.  The corresponding
swimmer's diffusion constants are predicted to get, respectively,
enhanced or suppressed upon increasing the model memory time.
Possible consequences of this effect on the interpretation of the
experimental data are discussed.

\end{abstract}
\maketitle

An artificial microswimmer enhances its diffusivity by harvesting
propulsion energy from its suspension fluid
\cite{Schweitzer,Reviews,Gompper}, as a result of some sort of
functional asymmetry of its own \cite{Sen_rev,Adjari,Sano}. The
particle thus propels itself with constant speed $v_0$, but keeps
changing direction due to both environmental fluctuations (and/or
spatial disorder) and the intrinsic randomness of the propulsion
mechanism itself.

For simplicity we restrict here our analysis to the
two-dimensional case of an overdamped swimmer with coordinates $x$
and $y$ in a fixed Cartesian frame, then its spatial diffusion is
described by the Langevin equations (LE) $$\dot x = v_0 \cos \phi
+\sqrt{D_0} \;\xi_x(t), ~~~\dot y = v_0 \sin \phi +\sqrt{D_0}\;
\xi_y(t),$$ where $D_0$ is the intensity of the thermal noise and
$\phi(t)$ denotes the instantaneous direction of the  propulsion
velocity with respect to the $x$ axis [see inset of Fig. 1(a)]. To
model the angular dynamics of the swimmer, two assumptions are
generally made in the current literature (with a few significant 
exceptions \cite{morelli,memory1,memory2} discussed below): (a) $v_0$ has a fixed
direction in the particle's reference frame, say, along some symmetry
axis, so that a change in the direction of the  propulsion velocity
implies a rotation of the swimmer; (b) Such rotations are assumed to
be mostly of thermal nature and, therefore, uncorrelated in time.
Accordingly, the swimmer's angular dynamics is modeled by a third LE,
\begin{equation} \label{phi0}
\dot \phi = \sqrt{D_\phi}\;\xi_\phi(t),
\end{equation}
where $D_\phi$ is the intensity of the rotational fluctuations. The
noises appearing in all three LEs are Gaussian, stationary, zero-mean
valued, and delta-correlated, that is $\langle \xi_i(t)
\xi_j(0)\rangle = 2\delta_{ij}\delta(t)$, with $i,j=x,y,\phi$. The
exact result \cite{Ghosh}, $\langle \cos \phi(t) \cos
\phi(0)\rangle=(1/2) \exp[-D_\phi t]$, obtained by combining the
general identity,
\begin{equation} \label{coscos}
\langle \cos \phi(t) \cos \phi(0)\rangle=(1/2) \exp[-\langle \Delta
\phi^2(t)\rangle/2],
\end{equation}
valid for any Gaussian process $\phi(t)$, and the mean square
displacement \cite{Risken}, $\langle \Delta \phi^2(t)\rangle=2D_\phi
t$, from the LE (\ref{phi0}), suggests to interpret the reciprocal of
$D_\phi$ as an angular diffusion time. Note that Eq. (\ref{phi0})
still describes a memoryless process.

The corresponding swimmer's spatial diffusion constant has also an
exact analytical expression \cite{EPJST},  that is
\begin{equation}  \label{Dstandard}
D\equiv \lim_{t\to \infty}\langle x(t)x(0)\rangle=\lim_{t\to
\infty}\langle y(t)y(0)\rangle=D_0+D_s,
\end{equation}
with $D_s=v_0^2/2D_\phi$. The quantities $v_0$ and $D_s$ are
experimentally accessible, so that $D_\phi$ is usually estimated from
the above model-dependent expression for $D_s$. Moreover, $D_0$ and
$D_\phi$ are often compared to assess their relationship as,
respectively, the translational and rotational constant associated
with a unique underlying diffusive mechanism
\cite{Bechinger,Lugli,Mijalkov,Kummel}. In this Communication we show
that releasing assumptions (a) and (b) itemized above impacts the
estimate of $D_\phi$ and, as a consequence, the interpretation of its
physical meaning.

{\it (1) Finite angular time correlation.} We expect that rotational
fluctuations of the swimmer and, therefore, random changes in the
direction of its velocity are caused to an appreciable extent by the
noisy nature of the  propulsion mechanism itself
\cite{Sen_rev,Adjari}. The complex interactions between the swimmer
and the active suspension fluid occur on finite spatio-temporal
scales \cite{Gompper,Sen_rev,Adjari}. The ensuing orientational
fluctuations of the swimmer, contrary to assumption (b), are thus
characterized by at least one finite relaxation rate, $\kappa_\phi$.
Stated otherwise, $\phi$ is more appropriately described by the
Ornstein-Uhlenbeck process,
\begin{equation} \label{phi1}
\ddot \phi = - \kappa_\phi \dot \phi +\kappa_\phi \sqrt{D_\phi}\;
\xi_\phi(t).
\end{equation}
The idea of adding an extra time scale, $\kappa_\phi^{-1}$, 
in the angular relaxation mechanism of an active Brownian particle is 
corroborated by the simulation work of Peruani and Morelli \cite{morelli}.
The second order LE (\ref{phi1}) reproduces the standard first-order stochastic
differential Eq. (\ref{phi0}) only in the limit of zero-correlation
time, i.e., for $\kappa_\phi\to \infty$. Accordingly, the angular
diffusion law now reads \cite{Risken}
\begin{equation} \label{Deltaphi1}
\langle \Delta \phi^2(t)\rangle = 2 D_\phi[t-(1-e^{-\kappa_\phi
t})/\kappa_\phi].
\end{equation}
The spatial diffusion constant follows immediately from Kubo's
formula \cite{EPJST},
\begin{equation} \label{Kubo}
D=D_0+\int_0^{\infty}C(t)dt,
\end{equation}
with $C(t)=v_0^2 \langle \cos \phi(t) \cos \phi(0)\rangle$. Since the
process in Eq. (\ref{phi1}) is Gaussian, $C(t)$ can be expressed in
terms of $\langle \Delta \phi^2(t)\rangle$, Eq. (\ref{Deltaphi1}), by
making use of Eq. (\ref{coscos}), which allows us to formally perform
the integration in Eq. (\ref{Kubo}); hence
\begin{equation} \label{D1total}
D=D_0+D_s\Gamma(D_\phi/\kappa_\phi)\sum_{m=0}^{\infty}\frac{(D_\phi/\kappa_\phi)^{m+1}}{\Gamma(m+1+D_\phi/\kappa_\phi)},
\end{equation}
where $\Gamma(x)$ denotes a gamma function. Two limits of this sum
can be calculated analytically:
\begin{equation} \label{D1min}
D\simeq D_0+D_s(1+D_\phi/\kappa_\phi),
\end{equation}
for $D_\phi/\kappa_\phi \ll 1$, and
\begin{equation} \label{D1max}
D\simeq D_0+D_s(\sqrt{(\pi/2)(D_\phi/\kappa_\phi)}-1),
\end{equation}
for $D_\phi/\kappa_\phi \gg 1$. Our analytical predictions compare
well with the simulation data obtained by numerically integrating the
model LEs \cite{Kloeden} [see Fig. 1(a)].

\begin{figure}[tp]
\centering
\includegraphics[width=0.5\textwidth]{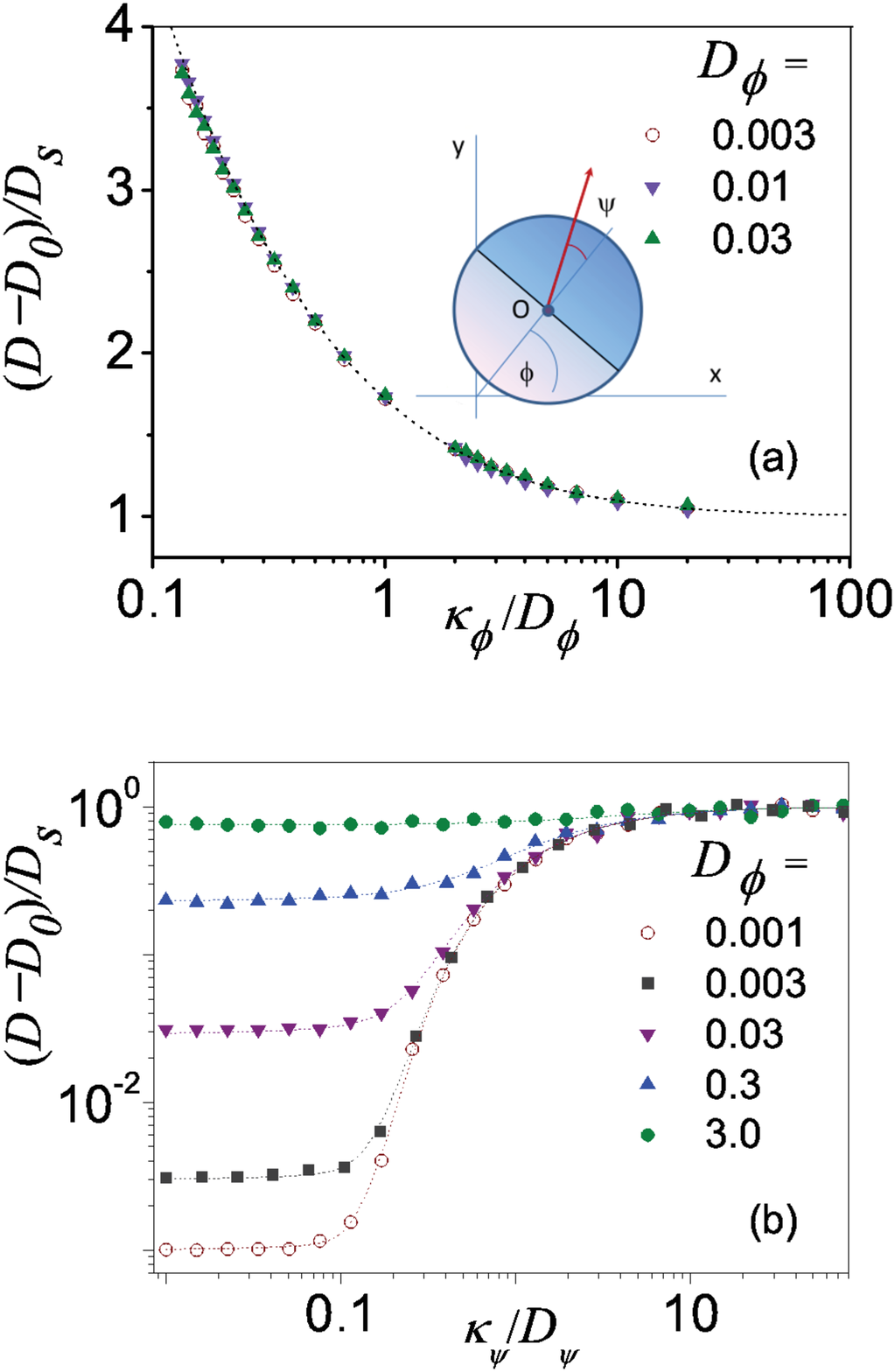}
%\vglue 0.5truecm \centering
%\includegraphics[width=7.3cm]{memFig1b}
\caption{(Color online) Diffusion constant versus angular relaxation
rate: (a) $D$ vs. $\kappa_\phi$ for model (1); (b) $D$ vs.
$\kappa_\psi$ for model (2) with $D_\psi=1$. In both panels $v_0=1$,
$D_0=0.01$ and $D_\phi$ is varied (see legends). The dashed curves in
(a) and (b) are the analytical predictions of Eqs. (\ref{D1total})
and (\ref{D2total}), respectively. The inset in (a) illustrates a
Janus particle with instantaneous propulsion velocity vector (red
arrow) directed at an angle $\psi$ with respect the symmetry axis of
angular coordinate $\phi$: in model (1) $\psi \equiv 0$; in model (2)
$\psi$ is the Ornstein-Uhlenbeck process described by the forth Eq.
(\ref{model2}). \label{F1}}
\end{figure}

{\it (2) Velocity fluctuations in the body frame.} Contrary to what
stipulated in assumption (a), the instantaneous direction of the
 propulsion velocity can fluctuate around its mean, represented
by the swimmer's axis of angular coordinate $\phi$. The resulting
swimmer's dynamics is thus modeled through a set of four LEs, namely
\begin{eqnarray} \label{model2}
\dot x &=& v_0 \cos (\phi+\psi) +\sqrt{D_0} \;\xi_x(t), \nonumber
\\\dot y &=& v_0
\sin (\phi + \psi) +\sqrt{D_0}\; \xi_y(t), \nonumber \\
\dot \phi &=& \sqrt{D_\phi}\;\xi_\phi (t), \nonumber \\
\dot \psi &=& -\kappa_\psi \psi +\sqrt{D_\psi}\; \xi_\psi (t),
\end{eqnarray}
where all noises are defined as above and the auxiliary angle $\psi$
represents the misalignment between the instantaneous  propulsion
velocity and its mean [see inset of Fig. 1(a)]. Here the restoring
constant $\kappa_\psi$ plays the role of a relaxation rate, whereas
the $\psi$ fluctuations have zero mean, $\langle \psi \rangle=0$, and
magnitude $\langle \psi^2 \rangle = D_\psi/\kappa_\psi$. For
$D_\psi/\kappa_\psi \ll 1$, the velocity fluctuations in the body
frame are suppressed and the standard model with $\psi\equiv 0$
recovered. The model of Eq. (\ref{model2}) exhibits normal diffusion
as always expected for realistic self-phoretic swimmers, as long as
one assumes observation times larger than $D_\phi^{-1}$ \cite{memory2}. In
this regard, we stress that the memory effects postulated here are intrinsic 
to the propulsion mechanism and should not mistaken for the additional inertial
and translational memory effects considered by Golestanian \cite{memory2}.

In order to apply Kubo's formula, Eq. (\ref{Kubo}), to evaluate the diffusion
constant, we need first to calculate the autocorrelation function
$C(t)=v_0^2 \langle \cos [\phi(t)+\psi(t)] \cos
[\phi(0)+\psi(0)]\rangle$. Simple algebraic manipulations yield
\begin{eqnarray} \label{C2}
C(t)=(v_0^2/2)e^{-\langle \psi^2\rangle}e^{-\langle \Delta
\phi^2(t)\rangle/2+\langle \psi^2\rangle e^{-\kappa_\psi t}},
\end{eqnarray}
with $\langle \Delta \phi^2(t) \rangle=2D_\phi t$.  Kubo's integral
can then be analytically calculated as a power series, i.e.,
\begin{equation} \label{D2total}
D=D_0+D_s e^{-D_\psi/\kappa_\psi}
\sum_{m=0}^{\infty}\frac{1}{m!}\frac{(D_\psi/\kappa_\psi)^{m+1}}{m(D_\psi/D_\phi)
+(D_\psi/\kappa_\psi)}.
\end{equation}
This expression is plotted against the simulation data in Fig. 1(b)
for different values of $D_\phi$.

In the realistic case when the velocity angular fluctuations in the
body frame are small, $D_\psi/\kappa_\psi \ll 1$, and their
relaxation time is much shorter than the body's characteristic
rotation time, i.e., for $\kappa_\psi/D_\phi\gg 1$, $D$ in Eq.
(\ref{D2total}) tends to
\begin{equation} \label{D2min}
D=D_0+D_s e^{-D_\psi/\kappa_\psi}.
\end{equation}
Vice versa, for a slowly relaxing $\psi(t)$, $\kappa_\psi/D_\phi\ll
1$, the spatial diffusion approaches the even smaller value
\begin{equation} \label{D2max}
D=D_0+D_s/(1+D_\psi/D_\phi).
\end{equation}
Both limits closely reproduce the numerical data of Fig. 1(b) within the
appropriate parameter range.

{\it Discussion.} We now compare models (1) and (2). In both cases
memory effects have been expressed in terms of a relaxation rate,
respectively, $\kappa_\phi$ and $\kappa_\psi$. In model (1) memory is
deemed as intrinsic to the nonequilibrium microscopic processes
responsible for the swimmer's propulsion -- for instance, a
shot-noise like sequence of {finite-time} pulses, or power-strokes,
associated with the chemical reactions catalyzed by the active tips
of the swimmer \cite{Sen_rev}. In model (2) we argued that the
direction of the shifts associated with such power-strokes may
fluctuate around its average orientation in the body frame as an
effect of the extended geometry of the swimmer's active. Of course, the 
above memory mechanisms might well operate simultaneously \cite{morelli}. This is
the case, for instance, of the reaction-driven swimmers of Ref. \cite{memory1},
where two sources of velocity fluctuations are singled out, namely,
the product particle density fluctuations and the randomness in the 
catalytic reaction that leads to the product particle release.

By inspecting Figs 1(a) and (b) we immediately recognize that finite
memory-time corrections to the spatial diffusion constant have
opposite sign: $D$ gets either enhanced or suppressed by decreasing
the corresponding model relaxation rate, namely, $\kappa_\phi$ in
Fig. 1(a), and $\kappa_\psi$ in Fig. 1(b). The physical
interpretation of these opposite behaviors is straightforward. In
model (1) lowering $\kappa_\phi$ means increasing the persistence
time of the propulsion mechanism, which is known to cause excess
diffusion \cite{Costantini}. On the contrary, in model (2) weakening
the restoring constant $\kappa_\psi$ favors the spatial reorientation
of the swimmer's kinematic velocity and, correspondingly, the
suppression of its spatial diffusion.

These remarks have a practical consequence on the interpretation of
the experimental data. Upon ignoring memory effects, one determines
$v_0$ and $D-D_0$ by direct measurements and extracts $D_\phi^{\rm
(exp)}$ from the identity, $D-D_0=v_0^2/2D_\phi^{\rm (exp)}$,
provided by the standard model, see Eq. (\ref{Dstandard}). However,
if we reconsider such a procedure in view of model (1), the computed
$D_\phi^{\rm (exp)}$ must differ from $D_\phi$. For instance, from
Eq. (\ref{D1min}) for $D_\phi/\kappa_\phi\ll 1$, $(D_\phi^{\rm
(exp)})^{-1}=D_\phi^{-1}+\kappa_\phi^{-1}$. The same conclusion holds
for model (2), where Eq. (\ref{D2min}) for $\kappa_\psi \gg
D_\phi,D_\psi$ implies that $D_\phi^{\rm (exp)}=D_\phi
e^{D_\psi/\kappa_\psi}$ and Eq. (\ref{D2max}) for  $\kappa_\psi \gg
D_\psi$ and $\kappa_\psi \ll D_\phi$ yields $D_\phi^{\rm
(exp)}=D_\phi+D_\psi$. In other words, $D_\phi^{\rm (exp)}$
systematically under- or over-estimates $D_\phi$, depending on which
model better reproduces the active swimmer's dynamics.

In the regime of short memory times the difference $|D_\phi^{\rm
(exp)}-D_\phi|$ is proportional to the ratios $D_\phi/\kappa_\phi$ in
model (1) and $D_\psi/\kappa_\psi$ in model (2). Being a measure of
the spatio-temporal structure of the  propulsion mechanism, the
parameters introduced in model (1) and (2), $\kappa_\phi$,
$\kappa_\psi$ and $D_\psi$, may vary with the physico-chemical
properties of the active fluid, the fabrication specifics (and
defects) of the active microswimmers, and their interactions with the
surrounding fluid. As a consequence, the estimated values of
$D_\phi^{\rm (exp)}$ also depend on all those factors. Therefore, the
measured quantity $D_\phi^{\rm (exp)}$ for an active swimmer cannot
be analyzed as the rotational counterpart of the translational
diffusion constant, $D_0$, if not after correcting, case by case, for
the memory effects peculiar of the propulsion mechanism actually at
work. Variations of the $D_\phi$ to $D_0$ ratio under different
swimmer's operation conditions have already been reported in the
earlier literature \cite{Sen_rev,Adjari,Lugli}.

{\it Concluding remarks.} In this Communication we emphasized the
dynamical role of rotational fluctuations. In a forthcoming
publication we will generalize both models (1) and (2) in various
ways. In model (1) the constant  propulsion speed, $v_0$, will be
explicitly derived from a {\it fluctuating} ``effective force'' to
mimic the microscopic discreteness of the propulsion mechanism.
However, since the average values of the such a propulsion force are
typically rather large in comparison to its standard deviation, no
significant contribution to the swimmer's diffusion is expected.
Model (2), instead, can be conveniently improved to account for
possible instability effects. Due to its functional asymmetry, the
center of mass and the center of the propulsion force acting upon a
swimmer, like the Janus particle in the inset of Fig. \ref{F1}(a),
may well rest a finite distance apart, say, along the symmetry axis.
As a consequence, the angular fluctuations of the  propulsion
velocity vector are associated with an additional instantaneous
torque. Although such a random torque has zero mean, it suffices to
further suppress active diffusion. Finally, for more asymmetric
geometries, where the fluctuating  propulsion velocity points in
average at an angle from the swimmer's axis, $\langle \psi \rangle
=\psi_0$, the ensuing nonzero average torque drives the swimmer along
spiraling trajectories. These generalizations of model (2) will allow
us to study the effects of chirality on active diffusion
\cite{Kummel,Takagi}.

\section*{Acknowledgements} We thank RIKEN's RICC for computational resources.
Y. Li is supported by the NSF China under grant No. 11505128.

\end{document}